%% file: conference_101719.tex
\documentclass[conference]{IEEEtran}
\IEEEoverridecommandlockouts
\usepackage{cite}
\usepackage{amsmath,amssymb,amsfonts}
\usepackage{listings}
\usepackage{algorithm}
\usepackage{algpseudocode}
\usepackage{graphicx}
\usepackage{subfigure}
\usepackage{cite}
\usepackage{comment}
\usepackage{multirow}
\usepackage{tcolorbox}
\usepackage{tabularx}
\usepackage{eqparbox}
\usepackage{adjustbox}
\usepackage{float}
\usepackage{xcolor}
\usepackage{mdwmath}
\usepackage{mdwtab}
\usepackage{eqparbox}
\usepackage{mdwmath}
\usepackage{url}
\usepackage{rotating}
{}
{}
{}
\usepackage{adjustbox}
\usepackage{placeins}
\usepackage{booktabs}

\usepackage{siunitx}
\usepackage{graphicx}
\usepackage{tikz}
\usetikzlibrary{positioning}
\usepackage{smartdiagram}

\definecolor{as_color}{rgb}{0.823, 0.411, 0.117}

\def\BibTeX{{\rm B\kern-.05em{\sc i\kern-.025em b}\kern-.08em
    T\kern-.1667em\lower.7ex\hbox{E}\kern-.125emX}}

\colorlet{punct}{red!60!black}
\definecolor{background}{HTML}{EEEEEE}
\definecolor{delim}{RGB}{20,105,176}
\colorlet{numb}{magenta!60!black}
\lstdefinelanguage{json}{
    basicstyle=\normalfont\ttfamily,
    numbers=left,
    numberstyle=\scriptsize,
    stepnumber=1,
    numbersep=8pt,
    showstringspaces=false,
    breaklines=true,
    frame=lines,
    backgroundcolor=\color{background}
    ,
    literate=
     *{0}{{{\color{numb}0}}}{1}
      {1}{{{\color{numb}1}}}{1}
      {2}{{{\color{numb}2}}}{1}
      {3}{{{\color{numb}3}}}{1}
      {4}{{{\color{numb}4}}}{1}
      {5}{{{\color{numb}5}}}{1}
      {6}{{{\color{numb}6}}}{1}
      {7}{{{\color{numb}7}}}{1}
      {8}{{{\color{numb}8}}}{1}
      {9}{{{\color{numb}9}}}{1}
      {:}{{{\color{punct}{:}}}}{1}
      {,}{{{\color{punct}{,}}}}{1}
      {\{}{{{\color{delim}{\{}}}}{1}
      {\}}{{{\color{delim}{\}}}}}{1}
      {[}{{{\color{delim}{[}}}}{1}
      {]}{{{\color{delim}{]}}}}{1},
}

\begin{document}

\title{Man-in-The-Middle Attacks and Defense in a Power System Cyber-Physical Testbed
}
\author{\IEEEauthorblockN{Patrick Wlazlo,
Abhijeet Sahu,
Zeyu Mao,
Hao Huang, 
Ana Goulart,\\
Katherine Davis, and
Saman Zonouz}
}

\maketitle
\thispagestyle{plain}
\pagestyle{plain}
\begin{abstract}
Man-in-The-Middle (MiTM) attacks present numerous threats to a smart grid. In a MiTM attack, an intruder embeds itself within a conversation between two devices to either eavesdrop or impersonate one of the devices, making it appear to be a normal exchange of information. Thus, the intruder can perform false data injection (FDI) and false command injection (FCI) attacks that can compromise power system operations, such as state estimation, economic dispatch, and automatic generation control (AGC). Very few researchers have focused on MiTM methods that are 
difficult to detect within a smart grid. To address this, we are designing and implementing multi-stage MiTM intrusions in an emulation-based cyber-physical power system testbed against a large-scale synthetic grid model to demonstrate how such attacks can cause physical contingencies such as misguided operation and false measurements. MiTM intrusions create FCI, FDI, and replay attacks in this synthetic power grid. This work enables stakeholders to defend against these stealthy attacks, and we present detection mechanisms that are developed using multiple alerts from intrusion detection systems and network monitoring tools. Our contribution will enable other smart grid security researchers and industry to develop further detection mechanisms for inconspicuous MiTM attacks.

\end{abstract}

\input{Content/1_introduction}

\input{Content/2_related_work}

\input{Content/3_background}

\input{Content/4_use_cases}
\input{Content/4_5_snort_ids}

\input{Content/5_results_analysis}

\input{Content/7_conclusion}

\section*{Acknowledgment}
This research is supported by the US Department of Energy's (DoE) Cybersecurity for Energy Delivery Systems program under award DE-OE0000895.

\bibliographystyle{IEEEtran}
\bibliography{reference.bib}


\end{document}

%% file: Content/1_introduction.tex
\section{Introduction}

The integration of information technology (IT) with industrial control systems (ICS) has revolutionized smart control of critical infrastructure systems such as energy, water, chemical, and transportation. Fast and accurate remote data collection and processing are helping to automate and optimize these sectors~\cite{fritz}. Initially, the integration of IT (cyber) and ICS (physical) systems focused on utility rather than data security. This lack of consideration of security in the design phase has staged serious security problems such as Stuxnet malware~\cite{stuxnet}, the Ukraine attacks~\cite{ukraine,ukraine2}, and the intrusion in the European Network of Transmission System Operators (ENTSO-E) in 2020~\cite{entso}. Due to the increasing awareness of such multi-stage, advanced concept-of-operations attacks, critical infrastructure stakeholders are focusing on the security of their networks, by following cyber-physical security policies that are unique to different sectors. These policies and decisions are governed by the threat models associated with the specific systems in each physical domain. In particular, the physical domain addressed in this paper is the energy sector, where we investigate man-in-the-middle (MiTM) attacks to an electrical utility’s supervisory control and data acquisition system (SCADA). 

A MiTM attack is one of the oldest forms of cyber intrusions, where a perpetrator positions itself in a conversation between two end points to either passively eavesdrop or actively impersonate one of the end points. MiTM attacks encompass different techniques, depending on the threat model. For example, Secure Socket Layer (SSL) hijacking is an attack where a \textit{man-in-the-middle} intercepts a request from client to server, then continues to establish an encrypted session between itself and the server, and a regular session between itself and the client, thus appearing to be a secure exchange between client and server. 

Similar to SSL hijacking, SCADA protocols such as Distributed Network Protocol-3 (DNP3), Modbus, and IEC-61850 can also be intercepted. If this happens, there is a potential to cause severe damage to energy systems, such as these scenarios: 

\begin{itemize}
    \item False data injection (FDI) can be performed that compromises state estimations~\cite{se_attack}, which affect economic dispatch~\cite{eco_dis}, generation scheduling, load forecasting, among others.
    \item False command injection (FCI) can be performed that can cause minor to major impacts, such as cascading failures and blackouts.
    \item Eavesdropping, where an intruder intercepts and reads packets, then uses this information to learn how the system operates. 
\end{itemize}


Furthermore, some MiTM attacks can be stealthy enough to evade conventional intrusion detection systems (IDS). First, the added latency caused by a man-in-the-middle intercepting packets may be difficult to detect in a wide area network (WAN). 
%
ICS data packets also have different delays depending on the application~\cite{wang}, and polling frequencies range from milliseconds to hours. 
Hence, delays that would occur due to MiTM can get masked based on the normal behavior of the system.
%
Second, most off-the-shelf security tools to detect MiTM attacks are designed for traditional internet applications and do not support ICS protocols. 
There are a lot of proprietary ICS protocols, and there are vendor-specific implementations of protocols like DNP3. Moreover, a testbed is needed to verify the impact of MiTM attacks and ways to mitigate them.  

Thus, our work addresses these gaps and challenges by focusing on MiTM attack methods that are 
difficult to detect within a smart grid and how to detect them. We are developing and implementing multi-stage MiTM intrusions in an emulation-based cyber-physical power system testbed against a large scale synthetic grid model to demonstrate how such attacks can cause physical contingencies such as misguided operation and falsified measurements. Then, to enable stakeholders to defend against these stealthy attacks, we also present detection mechanisms that are developed using multiple alerts from IDS and network monitoring tools, such as how we can correlate packet retransmission rates with alerts generated from the Snort IDS tool. 

The rest of this paper is organized into the following sections. In Section~\ref{related_works}, we evaluate previous papers on MiTM attacks performed within a cyber-physical testbed and their limitations.
Section~\ref{background} provides background on DNP3, MiTM attacks, and an overview of our testbed.
In Section~\ref{use_cases}, we present different methods for MiTM attacks on DNP3, such as binary operate and analog direct operate modifications, with detailed algorithms. We describe how an intrusion detection system can be configured to help detect two kinds of cyberattack in Section~\ref{detection}. In Section~\ref{results}, we present four MiTM attack use cases and analyze metrics that can be used to indirectly detect a MiTM attack. Conclusions and a review of our contributions are discussed in Section~\ref{conclusion}.

%% file: Content/2_related_work.tex
  \section{Related Work} 
\label{related_works}
There are several papers that describe MiTM attacks on energy systems that have cyber-physical system (CPS) testbeds and perform experiments on industrial control protocols. Their cyber-physical testbeds differ in terms of power and network simulators or emulators, amount and type of physical devices, and which energy system they model and in how much detail (e.g., power transmission or power generation). 
%
When comparing these papers, we observe certain limitations that are addressed by our work: i) they target small scale systems and support limited attack scenarios, ii) they do not include details on how the MiTM attack started, and iii) they do not show how to detect stealthy MiTM attacks.

In ~\cite{liu2015analyzing}, denial of service (DoS) attacks are performed in a CPS testbed which has a real-time digital simulator (RTDS) for power, network simulator-3 (NS-3) for communications, and devices such as phasor measurement units (PMUs) and phasor data concentrators (PDCs). They perform DoS attacks that increase delays in the communications links. The attack targets voltage stability monitoring and control in a transmission system. However, the method adopted in creating the DoS attack is not thoroughly presented in the paper.

The CPS testbed in~\cite{kezunovic2017use} uses RTDS, Opal-RT, and a WAN emulator to demonstrate cascading failures. The failures were caused by a coordinated data integrity attack that triggered the operation of a remedial action scheme (RAS). With false measurements, the MiTM attacks manipulated the automatic generation control (AGC) algorithm to take the wrong control action. Similarly, in~\cite{ashok2015experimental} AGC is targeted by MiTM attacks, where DNP3 packets carrying frequency and tie line flow measurements are modified using Scapy~\cite{biondi2010scapy} tools. Although the physical scenarios in these cases are demonstrated, the precursors of the MiTM attack are not explained clearly.

A Modbus-based MiTM attack on a CPS testbed is presented in~\cite{bo_karen}. The authors use Ettercap and LibModbus libraries to poison the address resolution protocol (ARP) cache and manipulate the Modbus packets, that affects the controller for a static volt-ampere reactive (VAR) compensator. To simulate a communication network, the authors use the Opnet simulator that provides system-in-the-loop (SITL) features to connect real devices to the simulator. Another work \cite{yang_2012} also uses existing libraries for performing a MiTM attack
on a grid-connected photovoltaic plant using the Metasploit framework. The testbed in~\cite{yang_2012} uses Schweitzer Engineering Laboratories (SEL) PMU hardware and the IEC 60870-5-103 protocol to demonstrate the attack scenario. However, in both cases, due to the limitations of the open source Metasploit and Ettercap frameworks,
only limited threat models are possible 
for the ICS protocols.

A multi-dimensional SCADA-specific IDS is presented in~\cite{yang_2017}. It detects MiTM attacks on IEC 61850 traffic in a real 500~kV substation. The MiTM attacks can easily be integrated in small test cases, but they have not been studied for large scale grids. 

Recent work on the Idaho CPS 
testbed~\cite{idaho_testbed} presents MiTM attacks on IEC C37.118, IEC 61850, and DNP3. However, the impact of the attacks on the power system is not presented, and the strategy adopted by incorporating the MiTM attacks is not clear. Another MiTM attack using DNP3 is presented in~\cite{darwish}, which also does not clearly illustrate the physical side threat model.

Authors in~\cite{fritz} present a MiTM attack on PMUs and PDCs by generating IEEE C37.118 packets using Wireshark. The use of Wireshark for creating a MiTM attack is unrealistic because it adds latency to the system that can be easily detected. 

In~\cite{DNP3_Eval} the time delay for ICS packets was studied. The authors found that for a normal relay's \emph{TRIP/CLOSE} command, the maximum tolerable delay was between 3~ms to 16~ms. The maximum delay for a human machine interface (HMI) workstation to receive updates was between 16~ms to 100~ms. This is a considerable short time frame for a MiTM intrusion to modify packets.


To complement these previous works, our contribution is to investigate MiTM threat scenarios in detail and how intercepted DNP3 packets can cause failures to the physical system. Using our own libraries to emulate the attacks, we implement and analyze use case scenarios in a CPS testbed that simulates a large scale synthetic electric grid based on the Texas footprint~\cite{synthetic,synthetic_comm}, where the attacker's stealthiness and its impact on five and ten simulated substations is evaluated. Such evaluations play a major role in exploring the threat space and proposing detection mechanisms.
Thus, we also present detection mechanisms that will enable other smart grid security researchers and industry stakeholders to detect similar MiTM attacks.


%% file: Content/3_background.tex
\section{Background}
\label{background}

This section gives a background on the implementation of multi-stage MiTM attacks for DNP3 in our emulation-based cyber-physical power system testbed. First, we present an overview of DNP3 and MiTM attack types. Our testbed, which allows us to model these MiTM threat and defense scenarios, is also explained in this section.


\subsection{DNP3}

DNP3~\cite{clarke2004practical} is a protocol used in SCADA systems for monitoring and controlling field devices. The protocol was released in 1993 for RS-485 serial links but has since been upgraded to work with TCP/IP networks. It can have multiple network setups using a master/outstation architecture. One example is a multi-drop network, where a DNP3 master communicates with more than one DNP3 outstation. Another example is a one-on-one network where a DNP3 master communicates with only one outstation.

There are three layers in the DNP3 protocol:

\begin{enumerate}

\item The \emph{data link layer} ensures the reliability of the physical link by detecting errors and duplicate frames. As shown in the example DNP3 packet in Fig.~\ref{fig:Generic_DNP3_PAcket_Structure}, the DNP3 header has 10 bytes, or octets, including two synchronization octets (\textbackslash x05 \textbackslash x64), followed by a frame length, data link control information field, and source and destination device addresses. At the end, there is a cyclic redundancy error (CRC) code to detect any bit errors in the header. 

\begin{figure}[b]
\centerline{\includegraphics[height=1.7 in,width=3.5 in]{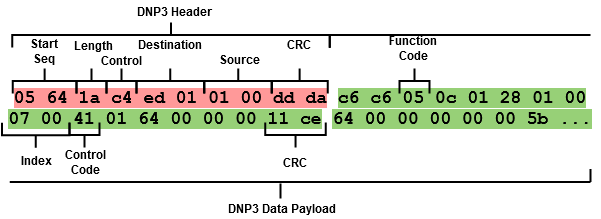}}
\caption{Hexadecimal representation of a DNP3 packet structure. 
}
\label{fig:Generic_DNP3_PAcket_Structure}
\end{figure}

\item The \emph{transport layer} supports fragmentation and reassembly of large application payloads. Using one octet, it stores FIR (1~bit), FIN (1~bit) and sequence number (6~bits), where FIR and FIN determine if the fragment is the first or the last fragment. The sequence number identifies each fragment so that they can be reassembled in the correct order before they are sent to the application layer. 

\item 
The \emph{application layer} provides services to the DNP3 user software so that DNP3 devices can send and receive messages. First, the application layer deals with DNP3 devices, known as DNP3 points, and then groups them according to their type: binary inputs ($BI$), binary outputs ($BO$), analog inputs ($AI$), analog outputs ($AO$), and counter input. Each group is identified by an index. Also, the application layer organizes static data and events into classes, where Class 0 means static data and Classes 1, 2, and 3 correspond to events with different priorities. Static data means the state of a DNP3 point, whereas an event means a change in the current state.
To indicate the purpose of the DNP3 message, the application layer header has a function code (FC) octet (Table~\ref{table:operation_codes}). There are two ways to send a command from the master to the outstation: the SELECT OPERATE where the master sends a select packet to a device in the outstation (FC:$03$), followed by the operation it should perform (FC:$04$); or the DIRECT OPERATE (FC:$05$), where one packet contains the device address with the operation it should perform. 
\end{enumerate}

Similar to other internet protocols, the DNP3 packet contains a header and a payload. The DNP3 payload has multiple \emph{data chunks}, consisting of 16-octet data blocks followed by a two-octet CRC to ensure each data block's integrity. Inside the payload, a function code is used to identify the operation the outstation should perform, as in Table~\ref{table:operation_codes}. The index will tell the outstation which device within the outstation the master is requesting the operation to be performed on, or retrieve data from. As shown in the sample packet in Fig.~\ref{fig:Generic_DNP3_PAcket_Structure}, we can see the hexadecimal representation of a binary DIRECT OPERATE DNP3 packet (FC:$05$).
This packet is indexed to close breaker seven in a substation of the Texas synthetic grid model. This is indicated by the index $07 00$, along with the $41$ control code to close the breaker.



\begin{table}[t]
\begin{tabular}{m{1.2cm}m{4.9cm}}
\cline{1-2}
 Function Code (Hex) & Operation   \\ \cline{1-2}
 \cline{1-2}
 0x00 &  Confirm    \\
 0x01 &  Read  \\ 
 0x02 &  Wire  \\ 
 0x03 &  Select \\  
 0x04 &  Operate \\  \cline{1-2}
 0x05 &  Direct Operate with Acknowledge \\  
 0x06 &  Direct Operate without Acknowledge \\    
 0x07 &  Freeze with Acknowledge \\  
 0x08 &  Immediate Freeze - No Acknowledge \\  
 0x09 &  Freeze and Clear with Acknowledge \\  \cline{1-2}
 0x10 & Freeze and Clear - No Acknowledge \\   
 0x13 & Cold Restart \\   
 0x14 & Enable Spontaneous Messages \\   
 0x15 & Disable Spontaneous Messages \\   
 0x16 & Assign Classes \\  \cline{1-2}
 0x17 & Delay Measurement \\  
 0x81 & Solicited Response \\ 
 0x82 & Unsolicited Response \\ \cline{1-2}
\end{tabular}
\caption{Function codes for DNP3 packets~\cite{dnp_func_codes}. 
}
\label{table:operation_codes}
\end{table}


As for confidentiality, DNP3 is a clear text -- unencrypted -- protocol with no inherent security mechanism~\cite{non_encrypt}. For this reason, the DNP3 protocol is susceptible to MiTM attacks, where an outsider can eavesdrop the communication between two nodes and modify the content of the packets. There have been numerous studies that try to incorporate encryption onto the DNP3 protocol using Transport Layer Security (TLS) encryption. However, this has not been widely adopted since maintaining time-sensitive public-key certificate server available for the DNP3 server and client requires costly upgrades to existing field equipment. 


\begin{figure*}[hbt!]
  \centering
\includegraphics[width=0.8\linewidth]{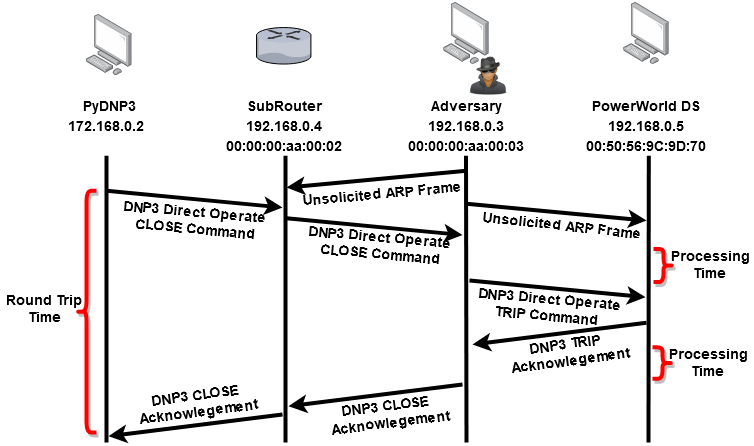}
  \caption{
  Timing diagram for ARP cache poisoning of the substation router ($SubRouter$) and $DNP3\;Outstation$ prior to the man-in-the-middle attack to modify the 
  $DIRECT\;OPERATE$ $CLOSE$ command.
  }
  \label{fig:flow_diagram}
\end{figure*}

\subsection{ARP Cache Poisoning}
The first step of the MiTM attack is for the adversary to impersonate a network device. This is done using address resolution protocol (ARP) spoofing, or ARP cache poisoning, where the adversary sends an unsolicited ARP messages to the targeted node. These messages are used to
link the adversary's hardware address -- or media access control (MAC) address -- with the internet protocol (IP) address of the targeted device. 


\begin{figure}[t]
  \centering
\includegraphics[width=1.0\linewidth]{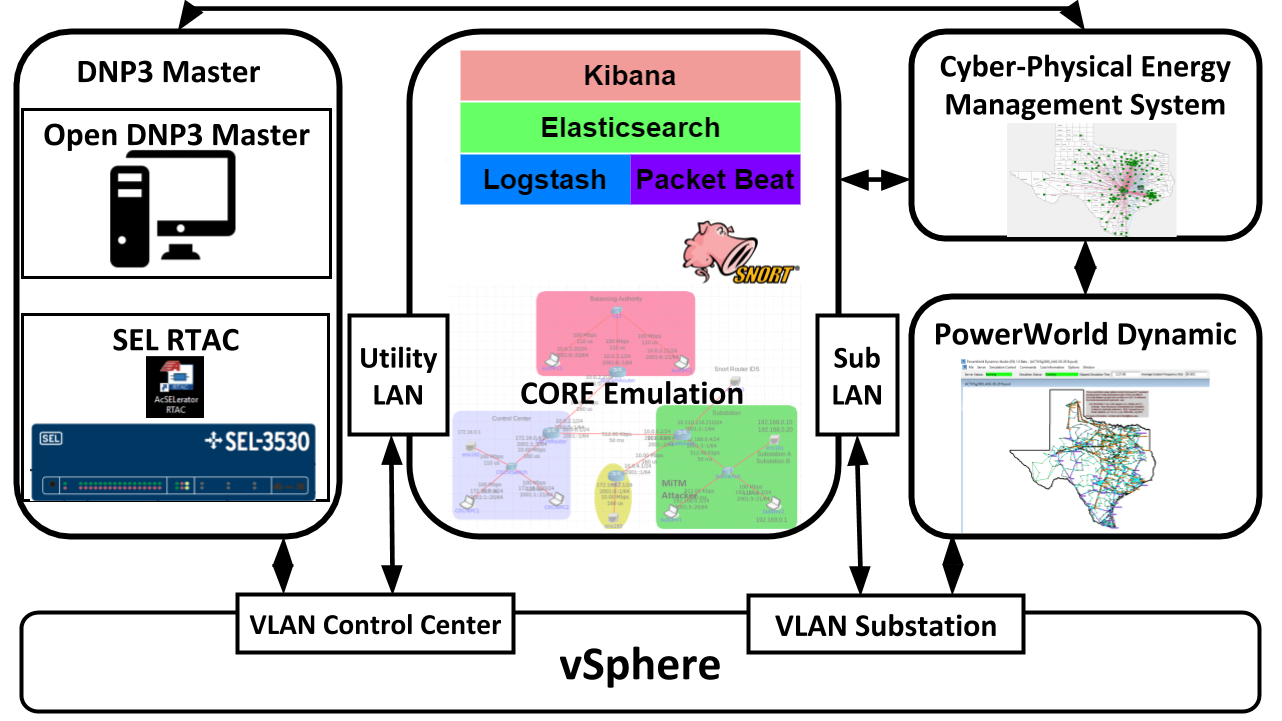}
  \caption{The RESLab emulation-based testbed architecture.}
  \label{fig:testbed_architecture}
\end{figure}

\begin{figure*}[!htb]
  \centering
\includegraphics[width=0.9\linewidth]{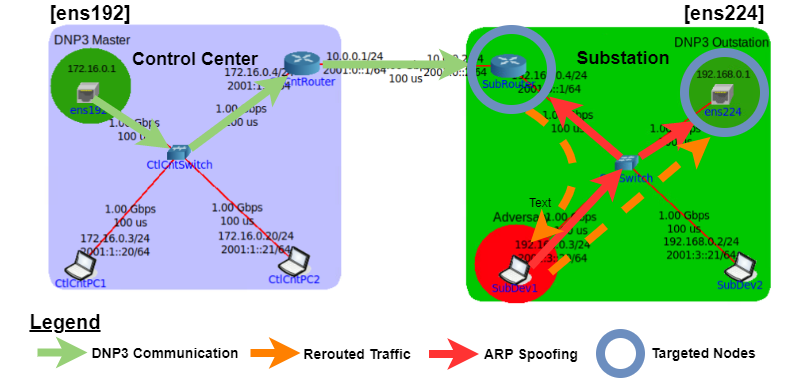}
  \caption{The RESLab testbed network topology emulated in CORE. The arrows show the data flows and virtual machine interconnections for our use cases (Section \ref{use_cases}).
  }
  \label{fig:core_topology}
\end{figure*}

As illustrated in Fig.~\ref{fig:flow_diagram}, the adversary sends an unsolicited ARP frame to the substation router telling the router to correlate the outstation's IP address with the adversary's MAC address. Thus, when the substation router needs to deliver a packet to the outstation, the router will 
instead forward it to the adversary.  Similarly, the adversary sends an unsolicited ARP packet to the outstation node so that it maps the router's IP address to the adversary's MAC address. After these unsolicited ARP packets, the adversary receives all packets sent between outstation and router. The adversary can now read or modify the packets' contents, before it forwards the packets to the correct device. The adversary in the example in Fig.~\ref{fig:flow_diagram} changes a DIRECT OPERATE command from \emph{CLOSE} to \emph{TRIP}, as well as a the response from \emph{TRIP} to \emph{CLOSE}.  As a result, the DNP3 communication channel remains open, and neither router nor outstation suspects that their packets are being intercepted.

The MiTM time diagram in Fig.~\ref{fig:flow_diagram} shows that the majority of the delay is at the adversary node, where the DIRECT OPERATE command is modified. This processing delay, or data injection delay, is the amount of time the adversary needs to filter the DNP3 packets, modify the packets' contents, and recalculate the DNP3 layers' CRC and TCP header checksum. A small portion of the delay time is due to the longer route the packet must travel after the ARP cache poisoning, since the packet is going through an extra node~--~the adversary.

\subsection{Integration of MiTM attacks in RESLab Testbed}
The MiTM attacks in our work are programmed to perform a staged intrusion, by trespassing into the broadcast domain of one substation's local area network (LAN). The trespassing could be a result of the adversary getting physical access to the substation site or by getting the credentials and remote access of one of the local devices. 

To simulate a substation LAN and control center, we use the RESLab testbed~\cite{testbed_architecture}, shown in Fig.~\ref{fig:testbed_architecture}, which is a CPS testbed comprised of the following components: 
\begin{itemize}
  \item \emph{Network Emulator} - the Common Open Research Emulator (CORE)~\cite{Ahrenholz2010core} emulates the communication network. CORE is a Linux-based application maintained by the U.S. Naval Research Laboratory, that uses FreeBSD jails configured as routers, switches, servers, and personal computers to create various emulated network nodes.
  \item \emph{DNP3 Master} - there are two different DNP3 masters in the RESLab testbed: the OpenDNP3 and the SEL-3530 real-time automation controller (RTAC). The OpenDNP3 is a DNP3 master application with a command line interface that is used to remotely operate outstation devices. The SEL-3530 RTAC is a cyber-physical component connected to the testbed, that in this instance is being used as a DNP3 master.
  \item \emph{DNP3 Outstation} - PowerWorld Dynamic Studio (PWDS) is a real-time simulation engine for high voltage power systems~\cite{powerworld}. In this paper, we use PWDS to simulate the synthetic Texas 2000-bus model~\cite{synthetic_comm} as our exemplar power system.
  \item \emph{Intrusion Detection System} - Snort~\cite{snort_cookbook} is being used in RESLab as the rule-based, open-source intrusion detection system (IDS). It is configured to generate alerts for ARP cache poisoning and FDI attacks.
  \item \emph{Storage and Visualization} - The Elasticsearch, Logstash, and Kibana (ELK)~\cite{elk} stack probes and stores all virtual and physical network interfaces' traffic, in addition to storing all Snort alerts generated during each use case. This data can be queried using Lucene queries to perform in depth visualization and cyber data correlation.
  \item \emph{Cyber-Physical Resilient Energy Systems (CYPRES) Application} - CYPRES aggregates information, i.e., from the cyber side CORE emulation environment, from the power side PWDS, as well as from the DNP3 masters regarding the communication status of DNP3 packets. All these data sets are then analyzed.
\end{itemize}

The RESLab testbed components are hosted in different virtual machines with the vSphere virtualization platform, as illustrated in Fig.~\ref{fig:testbed_architecture}.
In the middle, we have CORE which emulates a communication network that allows the DNP3 masters and outstation to interact. On the left side, CORE connects the utility control center, where there is a DNP3 master modeled using OpenDNP3 libraries and SEL-RTAC. On the right side, CORE connects the DNP3 outstations running in PWDS. 

The network topology of CORE is shown in detail in Fig.~\ref{fig:core_topology}, where the OpenDNP3 master and SEL-RTAC are connected to the CORE's network through virtual port \textit{ens192}. To emulate the outstations, the synthetic power system in PWDS is connected through virtual port \textit{ens224}. When the DNP3 communication link is not ARP cache poisoned, the traffic flows directly from \textit{ens192} to \textit{ens224}. However, when the adversary cache poisons the DNP3 communication link, all traffic between the control center and outstation passes through the adversary node, as shown by the dotted arrows.

%% file: Content/4_use_cases.tex
\section{MiTM Attacks on DNP3}
\label{use_cases}

\begin{table}[t]
\begin{tabular}{|c|c|} 
\hline
   Use case & Sequence of Function Call \\\hline
Use Case 1 & Alg.~\ref{alg:binary_control}  \\ \hline
Use Case 2 & Alg.~\ref{alg:analog_control}\\ \hline
Use Case 3 & Alg.~\ref{alg:sniff_measurement} $\rightarrow$  Alg.~\ref{alg:modify_measurement} $\rightarrow$ Alg.~\ref{alg:analog_control}\\ \hline
Use Case 4 &  Alg.~\ref{alg:sniff_measurement} $\rightarrow$  Alg.~\ref{alg:modify_measurement} $\rightarrow$ Alg.~\ref{alg:analog_control}$\rightarrow$ Alg.~\ref{alg:modify_measurement}\\ \hline
\end{tabular}
\vspace{0.05cm}
\caption{The sequence of function calls to the MiTM algorithms describe\\ in Section~\ref{use_cases} for each use case.}
  \label{fig:use_case_mapping}
\end{table}

In this section four different MiTM attack algorithms are described.
Each algorithm is used in a different sequence to generate the FDI and FCI use cases. 
 Table~\ref{fig:use_case_mapping} shows the order in which each algorithm is used for the four use cases.
 For example, in $Use\;Case\;3$, the adversary script calls Algorithms~\ref{alg:sniff_measurement},~\ref{alg:modify_measurement}, and~\ref{alg:analog_control}. The function of each algorithm is described in the proceeding paragraphs.




\subsection{Scapy MiTM Script}
The MiTM attack scripts are programmed using the Scapy Project Python wrapper \cite{biondi2010scapy}. Scapy is a powerful library that can modify frames and/or packets in real-time. The library has many built-in packet dissectors for applications using either user datagram protocol (UDP) or transmission control protocol (TCP). However, DNP3 is not one of the supported TCP applications. Fortunately, \cite{rodofile2015real} introduces a Scapy extension for DNP3, which we use to dissect and filter DNP3 packets.

Here is how it works. Scapy reads all traffic sent to the adversary once the route is ARP cache-poisoned. Then, the DNP3 extension along with Scapy's native libraries filters traffic based on IP, TCP, and DNP3 header information. For instance, if the function code of a captured DNP3 packet is (FC:$05$), it indicates an analog or a binary DIRECT OPERATE command. Thus, the adversary needs to have its value modified before it forwards the packet to the original destination. If the TCP packet does not have a function code, it is not a DNP3 packet and the adversary only forwards it to the destination. 

This process is not as straightforward as it may seem. First, the total process must be optimized to take the least amount of processing time as possible. Second, the DNP3 traffic is filtered by function code and then by control code to determine if the payload is an analog or binary command. Third, in order to keep the operator unaware that the wrong command has been sent to the outstation, the acknowledgement number of each DNP3 command packet is used to filter the appropriate response packet, that is changed to match the original command sent by the control center.

\subsection{Binary Operate}
Substation breakers are represented by binary points ($BI$ or $BO$) that can have their states updated remotely by DNP3 binary DIRECT OPERATE packets. Each point can either be opened (tripped) or closed. The \emph{TRIP} action will disconnect or de-energize a line, a \emph{CLOSE} will energize an open a line. The MiTM script inverts the binary command. In other words, a \emph{TRIP} command is forced to a \emph{CLOSE} command, or vice-versa.

Algorithm~\ref{alg:binary_control} describes the process to invert a binary DIRECT OPERATE packet. After a received  packet ($recv\_pkt$) is identified as a DNP3 binary DIRECT OPERATE packet, its TCP header checksum is removed, because Scapy automatically recalculates the TCP header checksum if there is not one detected when forwarding the frame. The $recv\_pkt$'s acknowledgement number is stored as $binary\_operate\_ack$, so the response packet can be changed to the original binary value. Then, the DNP3 header is stored as $dnp\_header$. The same is done for the packet's payload or $dnp\_pl$. Next, the payload is bisected around the \textit{control code}, as in Fig.~\ref{fig:Generic_DNP3_PAcket_Structure}. The first half of the payload is stored under the $dnp\_front$. The second half is stored under the $dnp\_end$. If the packet's control code is a \emph{CLOSE} command (`41'), it is modified to be a \emph{TRIP} command (`81'), or vice-versa. 

\begin{algorithm}
  \caption{MiTM attack on binary control commands}\label{alg:binary_control}
	\begin{algorithmic}[h]
	\Function{$modify\_binary\_direct\_operate$}{$recv\_pkt$}
      \State {$binary\_operate\_ack$ = $recv\_pkt[TCP].ack$}
      \State {$mod\_pkt$ = $recv\_pkt[TCP]$}
      \State {Delete $mod\_pkt.checksum$}
      \State {$dnp\_header$ = $mod\_pkt.pl[:dnp\_hdr\_size]$}
      \State {$dnp\_pl$ = $mod\_pkt.pl[dnp\_hdr\_size:]$}
      \State {$dnp\_front$ = $mod\_pkt.pl[dnp\_hdr\_size:bin\_loc]$}
      \State {$dnp\_end$ = $mod\_pkt.pl[bin\_loc + 1:]$}
      \State{$dnp\_mid$ = $mod\_pkt.pl[bin\_loc]$}
      \If{$dnp\_mid$ == $b'\textbackslash x41'$}       
      $dnp\_mid$ = $b'\textbackslash x81'$
      \Else \;
         $dnp\_mid$ = $b'\textbackslash x41'$
      \EndIf
      \State {$merged\_pl = Join[dnp\_front, dnp\_mid, dnp\_end]$}
      \State {$pl\_with\_crc = update\_crc\_payload(merged\_pl)$}
      \State {$mod\_pkt.pl$ = $Join[dnp\_header, pl\_with\_crc]$}
      \State {$mod\_pkt$ = $send\_to\_outstation(mod\_pkt)$}\\
      \Return {$mod\_pkt, binary\_operate\_ack$}
    \EndFunction
	\end{algorithmic} 
\end{algorithm}

%
Then, the DNP3 payload is reassembled by joining the $dnp\_front$, $dnp\_mid$, and $dnp\_end$ together as $merged\_pl$. The reassembled payload is passed to the $update\_crc\_payload$ function, which comes from the DNP3 Scapy extension~\cite{rodofile2015real}.
Finally, the MAC address in the frame's header is updated to the MAC address of the outstation, and the adversary forwards the frame to the outstation. 

\subsection{Analog Direct Operate}
The setpoints of generators and other controls are represented by analog points ($AI$ or $AO$) in the DNP3 protocol. Each setpoint can be varied by the DNP3 master. In the MiTM script, any analog DIRECT OPERATE setpoint is forced to a lower value.
The lower value ramps down the generator without tripping it. 
Algorithm~\ref{alg:binary_control} inverts the control code for a binary DIRECT OPERATE command. Algorithm~\ref{alg:analog_control} changes analog values instead of binary values. The main difference between the two algorithms is $dnp\_mid$ which is an analog value that is modified in Algorithm~\ref{alg:analog_control}. The analog value is a four-octet float value that is encoded with a one-octet control status. In the $update\_new\_val()$ function, the analog value in the original $recv\_pkt$ is changed to a forged 
value. Once the forged value is placed in the correct position, the DNP3 payload CRC, TCP header checksum, and the MAC address are updated before the packet is forwarded to the outstation.

\begin{algorithm}
  \caption{MiTM attack on analog control commands}\label{alg:analog_control}
	\begin{algorithmic}[1]
	\Function{$modify\_analog\_direct\_operate$}{$recv\_pkt$}
      \State {$analog\_operate\_ack$ = $recv\_pkt[TCP].ack$}
      \State {$mod\_pkt$ = $recv\_pkt[TCP]$}
      \State {Delete $mod\_pkt.checksum$}
      \State {$dnp\_header$ = $mod\_pkt.pl[:dnp\_hdr\_size]$}
      \State {$dnp\_pl$ = $mod\_pkt.pl[dnp\_hdr\_size:]$}
      \State {$dnp\_front$ = $mod\_pkt.pl[dnp\_hdr\_size:alg\_loc]$}
      \State {$dnp\_end$ = $mod\_pkt.pl[anlg\_loc+5:]$}
      \State{$dnp\_mid$ = $mod\_pkt.pl[anlg\_loc+1: anlg\_loc+5]$}
      \State{$dnp\_mid$ = $update\_new\_val(dnp\_mid)$}
      \State {$merged\_pl = Join[dnp\_front, dnp\_mid, dnp\_end]$}
      \State {$pl\_with\_crc = update\_crc\_payload(merged\_pl)$}
      \State {$mod\_pkt.pl$ = $Join[dnp\_header, pl\_with\_crc]$}
      \State {$mod\_pkt$ = $send\_to\_outstation(mod\_pkt)$}\\
      \Return {$mod\_pkt, analog\_operate\_ack$}
    \EndFunction
	\end{algorithmic}
\end{algorithm}

\subsection{Polled Measurement Sniff and Store}
To target the intended DNP3 packet, the MiTM script must first sniff through all the network traffic between the substation gateway and PWDS. Then, each packet the MiTM script receives is filtered by its function code. For every fifth packet with a DNP3 function code of `81', the analog and binary data value is stored in the adversary's machine. Not every read response packet is stored since the processing time for these packets is relatively high and would lead to more retransmitted packets.

Algorithm~\ref{alg:sniff_measurement} describes how the read response packets are stored in a $dnpDatabase$. 
A DNP3 response payload for poll request consists of collection of the DNP3 points stored in the \emph{datachunks} with the $chunk\_size$ of 18 bytes (16 bytes of payload and two bytes of CRC).  
First, the packet data $dnp\_pl$ is checked to see if it contains one or more \emph{datachunks}. For the payload with at least one \emph{datachunk}, each $dnp\_chunk$'s CRC is removed and its contents are concatenated into contiguous bytes of $BI$, $AI$, $BO$, and $AO$ points. Then, based on the header information $bi\_hdr$, $ai\_hdr$, $bo\_hdr$, $ao\_hdr$, the number of DNP3 points under each category $bi\_count$, $ai\_count$, $bo\_count$, $ao\_count$ is extracted. The information such as the $pointIndex$, $value$, $chunkIndex$, $pointType$ of each DNP3 point is stored in a $dnpDatabase$ classified by the source address of the packet, which is unique to each outstation number.
The $pointIndex$ stores the actual DNP3 index. The $value$ stores the value associated with the DNP3 point. The $pointType$ indicates the type of the DNP3 point. The purpose of storing $chunkIndex$ is to identify the location of the DNP3 point in the \emph{datachunks}. These attributes are further used by the intruder in Algorithm~\ref{alg:modify_measurement} to modify the measurement in a specific location, which results in a faster modification of the DNP3 payload.

\begin{algorithm}
  \caption{MiTM attack on sniffing measurements}\label{alg:sniff_measurement}
	\begin{algorithmic}[1]
	\Function{$sniff\_read\_response$}{$recv\_pkt, outstation$}
      \State {$mod\_pkt$ = $recv\_pkt[TCP]$}
      \State {Delete $mod\_pkt.checksum$}
      \State {$dnp\_header$ = $mod\_pkt.pl[:dnp\_hdr\_size]$}
      \State {$dnp\_pl$ = $mod\_pkt.pl[dnp\_hdr\_size:]$}
      \State {$chunk\_size$ = 18}
      \State {$dnp\_chunks$ = $len(dnp\_pl)/chunk\_size$}
      \If {$dnp\_chunks == 0$} 
       \State $send\_to\_master(mod\_pkt)$
       \Return $mod\_pkt$
      \EndIf
      \State {Store each data chunks in $dnp\_chunks\_pl$}
      \State {$dnp\_pl\_without\_crc$ = Remove 2 bytes of CRC from each $dnp\_chunks\_pl$}
      
      \State {$dnp\_reassembled$ = $Join(dnp\_pl\_without\_crc)$}
      \State {Obtain $bi\_hdr$, $bi\_pl$, $ai\_hdr$, $ai\_pl$,$bo\_hdr$, $bo\_pl$,$ao\_hdr$, $ao\_pl$ using $bi\_count$,$ai\_count$, $bo\_count$, $ao\_count$}
      
      \State {Obtain $start$ and $end$ index for $BI, AI , BO, AO$ group types}
      \State {Store $DNP3Points$ for each point types}
      \State {Create $dnpDatabase$ for $outstation$ using each point in $DNP3Points$ for $BI, AI , BO, AO$ types}
      
      \Return {$dnpDatabase$}
      
    \EndFunction
	\end{algorithmic}
\end{algorithm}

\subsection{Polled Measurements Modification}
Periodically, the master polls each outstation that is connected to it for updates on the binary and analog points. In the RESLab testbed, the polling interval varies from 30 to 60~s. When polled, an outstation responds with a list of all the binary and analog points housed within that outstation. There are multiple ways this data can be manipulated. For instance, the poll measurements can be spoofed to a wrong value, causing the operator to send the incorrect command to the outstation. 
Or these updates can be forged so the operator will choose not to send a command to an outstation, when he/she should.  

\begin{algorithm}
  \caption{MiTM attack on modifying measurements}\label{alg:modify_measurement}
	\begin{algorithmic}[1]
	\Function{$modify\_read\_response$}{$dnpDatabase$,$ModPoints$}
	  	\State {$mod\_pkt$ = $recv\_pkt[TCP]$}
	    \State {Delete $mod\_pkt.checksum$}
        \For {$dnp3Point$ in $ModPoints$}
            \State {$dnp\_header$ = $mod\_pkt.pl[:dnp\_hdr\_size]$}
            \State {$dnp\_pl$ = $mod\_pkt.pl[dnp\_hdr\_size:]$}
            \If {$dnp3Point.pointType$ is $BI \lor BO$ }
                \State {$bin\_loc$ = using $dnp3Point.chunkIndex$ and $dnp3Point.pointIndex$ }
                \State {Follow steps 7 to 15 in Alg.~\ref{alg:binary_control}}
            \EndIf
            \If {$dnp3Point.pointType$ is $AI \lor AO$ }
                \State {$anlg\_loc$ = using $dnp3Point.chunkIndex$ and $dnp3Point.pointIndex$ }
                \If{$len(dnp3Point.chunkIndex) > 1 $ }
                 \State Process the $dnp3Point.chunkIndex[0]$ and $dnp3Point.chunkIndex[1]$ separately
                \EndIf
                \If{$len(dnp3Point.chunkIndex) == 1 $ }
                    \State {Follow steps 7 to 14 in Alg.~\ref{alg:analog_control}}
                \EndIf
            \EndIf
        \EndFor
      \State {$mod\_pkt$ = $send\_to\_master(mod\_pkt)$}\\
      \Return {$mod\_pkt$}
    \EndFunction
	\end{algorithmic}
\end{algorithm}

The $dnpDatabase$ containing all the datapoints captured in Algorithm~\ref{alg:sniff_measurement} is passed to the $modify\_read\_response()$ function, shown in Algorithm.~\ref{alg:modify_measurement}. A list of $BI$, $BO$, $AI$, and $AO$ points that the MiTM attack intends to use are contained in the variable $ModPoints$ that is passed to the function. Each point listed in $ModPoints$ is modified in the $dnpDatabase$, and the CRC for each \emph{datachunk} is updated. $BI$ and $BO$ points have only one-octet and therefore can be contained in one \emph{datachunk} (lines 7-10 in Algorithm~\ref{alg:modify_measurement}). However, $AI$ and $AO$ are five octets long and can be split between two datachunks (lines 11-19 in Algorithm~\ref{alg:modify_measurement}). Each \emph{datachunk} that is modified must have its CRC recalculated for the master to accept the packet.

\subsection{Acknowledgements Modification}
After a binary or analog DIRECT OPERATE packet is received at an outstation, a DNP3 acknowledgement packet is returned stating the action was performed. When the MiTM script changes the binary or analog command or value to perform an incorrect action, the outstation's response is a telltale sign that the DNP3 communication channel is compromised. For the MiTM to remain unnoticed by the control center's operator, the DNP3 acknowledgement from the outstation must be modified.
When a binary \emph{CLOSE} command is sent by the operator, the MiTM scrip changes the command to a binary \emph{TRIP} and then forwards it to the outstation. Correspondingly, in the acknowledgement, the outstation sends a binary response stating that the breaker is opening. The intruder then modifies 
the true response into a binary \emph{CLOSE} response and forwards it to the operator. This leaves the DNP3 master unaware that the wrong action has been sent to the outstation.

Algorithm~\ref{alg:binary_control}, which is used to modify the binary DIRECT OPERATE packet, modifies the binary operate response packet. Similarly, Algorithm~\ref{alg:analog_control} modifies the analog operate response packet.  

%% file: Content/4_5_snort_ids.tex
\section{Snort Configuration for MiTM Attack}
\label{detection}

It is essential for utility companies to monitor and secure their networks from various forms of cyber threats. Generally, this comes in the form of an IDS that can detect vulnerabilities in a network and generate alarms. 
Snort is a open-source IDS~\cite{snort_cookbook} that can be configured to dissect Ethernet packets to monitor for a variety of attacks. Each type of attack has a pre-processor which can be enabled in the Snort's configuration file. Then, rules based on the data the pre-processors collects are created to generate alerts. The alerts can be displayed in real-time or saved to a file.

During each trial, Snort is running in the substation gateway or \textit{SubRouter} (IP: $192.168.0.4$), shown in Fig.~\ref{fig:core_topology}. The Snort ARP and DNP3 pre-processors are used. In the Snort configuration file, the MAC addresses of the $SubRouter$ and $DNP3\;Outstation$ (IP: $192.168.0.5$)
are white-listed (Listing~\ref{preproc}),
where a list of known IP addresses and their MAC addresses within a LAN is maintained by Snort~\cite{white_list}.
This allows it to detect if the MAC address has changed from the listed IP address, which indicates an attempt to poison the ARP table of the router. 
The DNP3 pre-processor (Listing~\ref{dnp_preproc}) detects a DNP3 packet and checks if the CRC is correct; if not, an alert is generated. 
The pre-processor is configured to detect when a DNP3 DIRECT OPERATE packet is sent.


\lstset{
  breaklines=true,
  postbreak=\mbox{\textcolor{red}{$\hookrightarrow$}\space}
}
\begin{lstlisting}[language={C},caption={ARP pre-processor enabled in Snort configuration file},label={preproc}]
preprocessor arpspoof_detect_host: 192.168.0.4 00:00:00:aa:00:02
preprocessor arpspoof_detect_host: 192.168.0.5 00:50:56:9c:9d:70
\end{lstlisting}


\begin{lstlisting}[language=C,caption={DNP3 pre-processor enabled in Snort configuration file },label={dnp_preproc}]
preprocessor dnp3: ports {20000} \
    memcap 262144 \
    check_crc
\end{lstlisting}

With the pre-processors enabled, custom rules are created. Rules are added to generate logs that can be used to alert the operator. The first three alerts, $R1$, $R2$, and $R3$, indicate an ARP cache poisoning. The next two alerts, $R4$ and $R5$, notify when a DNP3 DIRECT OPERATE packet is sent to the outstation.

\begin{lstlisting}[language=C,caption={ARP and DNP3 specific alerts configured in Snort},label={snort_rules}]
R1 alert ( msg: "ARPSPOOF_ETHERFRAME_ARP_MISMATCH_SRC"; sid: 2; gid: 112; rev: 1; metadata: rule-type preproc ; classtype:bad-unknown; )
R2 alert ( msg: "ARPSPOOF_ETHERFRAME_ARP_MISMATCH_DST"; sid: 3; gid: 112; rev: 1; metadata: rule-type preproc ; classtype:bad-unknown; )
R3 alert ( msg: "ARPSPOOF_ARP_CACHE_OVERWRITE_ATTACK"; sid: 4; gid: 112; rev: 1; metadata: rule-type preproc ; classtype:bad-unknown; )
R4 alert tcp $EXTERNAL_NET any -> 192.168.0.5 20000 (msg:"DNP3 Snort DIRECT OPERATE"; flow:established, to_server; dnp3_func:direct_operate; sid:123000;)
R5 alert tcp $EXTERNAL_NET any -> 192.168.0.5 20000 (msg:"DNP3 Snort OPERATE"; flow:established, to_server; dnp3_func:operate; sid:123002;)
\end{lstlisting}



%% file: Content/5_results_analysis.tex
\section{Results and Analysis}
\label{results}


This section briefly describes the experiment setup in the RESLab testbed, then presents the results of four MiTM use cases. In addition, we show how 
Snort alerts can be used to detect the MiTM attack.
Snort is operating in a Network Intrusion Detection System (NIDS) mode at the substation router, protecting the substation's LAN.

\subsection{Experimental Setup}


As shown in RESLab testbed in Fig.~\ref{fig:testbed_architecture}, the DNP3 master and the SEL RTAC are connected through vSphere's control center virtual local area network (VLAN) to the CORE emulator. 
The DNP3 outstations, modeled in PWDS, are connected through vSphere's substation VLAN to the CORE network. 
%
In most known cyber attacks on an ICS network, the intruder had to perform multi-stage intrusions to reach the targeted grid components. Since this work focuses on the dynamics of MiTM attacks, the prior stages do not play a major role. Hence, we assume the intruder, after a reconnaissance stage, has remote access to one of the computer nodes in the substation LAN, which in this instance is the adversary node.  


\subsection{Evaluation Metrics}
The strength of the MiTM attack is determined by analyzing the average round trip time (RTT), retransmission rate, and average processing time of DNP3 packets, as described below.  

\begin{enumerate}
    \item \textbf{Retransmission Rate}:
    When a packet is sent, the sender starts a variable-length retransmission timer, and waits for the acknowledgement. If it does not receive an acknowledgement before the timer expires, the sender assumes the packet is lost and retransmits it. During the MiTM attack, the number of retransmissions increases, because packets may not be successfully forwarded to the outstation and the DNP3 master may not receive the acknowledgement. This may also happen if the adversary cannot forward the acknowledgement it receives from the outstation. 
    Since the duration of each use case varies, retransmission rate is used as a metric instead of the number of retransmissions. The retransmission rate $R_R$ is computed using Eq.~\ref{eq:retransmission_rate}, 
    \begin{equation}
    R_R = N_R/T_R 
    \label{eq:retransmission_rate}
    \end{equation}
    
    where $N_R$ is the number of retransmitted packets during the MiTM, and $T_R$ is the time interval between the first and the last retransmitted packet in seconds. 

    \item \textbf{Average Round Trip Time (RTT)}: The RTT can be seen in the time diagram in Fig.~\ref{fig:flow_diagram}. It includes the network's propagation delay due to the distance between nodes, the added transmission delays as the packet travels through the adversary node, and the processing time the adversary takes to modify the commands and response. Hence, we evaluate the impact of a MiTM attack on the RTT. 
    
    \item \textbf{Processing time}: The processing time depends on the type of DNP3 traffic the intruder modifies. The processing time for modifying outstation polled responses can vary based on the outstation data that is polled. The outstation's read response depends on the number of DNP3 points housed at a particular outstation.

\end{enumerate}

The retransmission rate and average RTT are extrapolated by analyzing Wireshark packet captures (PCAP) data from the \textit{SubRouter}'s network interface. The processing delay is automatically calculated by the MiTM attack script. 



\subsection{Modifying Measurements and Commands}
\label{usecases}
The objective of the intruder is to 
disrupt grid operations. 
Details on the sequence of actions that create the FCI and FDI attacks and how they impact the physical components of the power system are presented in detail in~\cite{testbed_architecture}. Here in this paper we focus on the impact of the attacks on the communications network, or cyber telemetry. These are our four use cases:

\subsubsection{Use Case 1: Branch Control Modifications. \:} 
Each binary DIRECT OPERATE command is changed from a \textit{CLOSE} to a \textit{TRIP} command, with any other traffic simply forwarded. The change in the binary operate command introduces some processing delay, which may cause the packet to be retransmitted.

\subsubsection{Use Case 2: Generator Set-Point Modification. \:}
When the MiTM script is running, the analog point for the generator is set to a lower value, in some cases 20~MW, which will decrease the generator setpoint from its current value down to 20~MW.

\subsubsection{Use Case 3: Measurement and Status Modification. \:}
Use Case 3 is a combination of FCI and FDI attacks. After each polling interval, the DNP3 master will send a read request packet to each outstation, which then sends a read response packet back to the master. This read response is filled with the all the binary input, analog input, binary output, and analog output DNP3 points. Next, analog input points in the read response packet are changed to a lower value lower of 20~MW or 0~MW. The operator controlling the DNP3 master is then forced to send an analog DIRECT OPERATE command to bring the generators back to their original loaded set points. However, when the operator sends this original set point value to the generator, the MiTM script is programmed to change the setpoint to 20~MW or 0~MW. 

\subsubsection{Use Case 4: Measurement and Status Modification. \:}
The adversary first follows the steps of Use Case~3, then modifies the read response packet of the preceding packets, based on the actual set point given by the master. Thus, the master is unaware of the contingency created.

\subsection{Use Cases Implementation}
For each use case, we alter the polling intervals and the number of polled DNP3 outstations. The polling intervals tested were 30 and 60~s, while the number of polled DNP3 outstations were five and ten. For instance, the scenario $UC1\;10\;OS\; 30$ means that we implemented $Use\;Case\;1$ with ten outstations and a polling interval of 30~s. 
In each scenario, the normal operation is conducted first without the MiTM attack. Then, the operation is conducted again with the attack to analyze its impact. Finally, the attack is stopped and the network restored.

The main reason for choosing polling intervals of 30 and 60~s is that most DNP3 masters have polling rates of 30~s, 1~min, or 5~min, with a maximum of 15~min.
A polling interval of more than two minutes has little impact on attack strength
because the adversary processing time is less than 60 to 70~ms (see Section \ref{subsec:processingtime}).

Similarly, we choose outstation numbers of five and ten since our objective is to study the communication dynamics of an impacted outstation, 
and how the number of outstations becomes a limitation on the 
attack success probability. The numbers of five and ten coincide with our use cases in the Texas 2000-bus model where each utility control center communicates with at least two and at most 25 substations. 
Because the RESLab testbed uses CORE, which is an emulator and not a simulator, there are practical limitations to the number of substations that can be modified by the MiTM script. This number in our testbed is about 50 substations; however, this depends on the amount of memory and capacity of the network interface card's buffer that is allocated to CORE. Since CORE is an emulator, it demonstrates more realistically the physical limitations an actual adversary would have to experience in order to create a successful MiTM attack. 

\subsection{Impact of Polling Rates and Number of Outstations on Retransmission}
\begin{figure}[b]
  \centering
\includegraphics[width=1\linewidth]{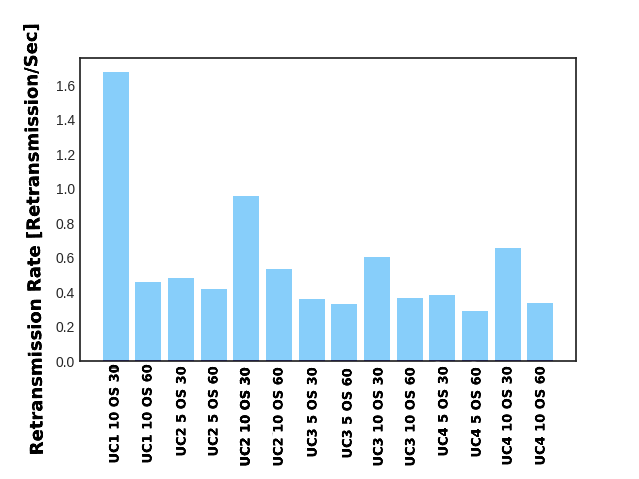}
  \caption{Retransmission rate of DNP3 traffic for each scenario.}
  \label{fig:retransmission_rate}
\end{figure}

 High polling rates, or low polling intervals, result in packet losses during an attack due to the limitations of an adversary's resources to process all the command and response DNP3 traffic. Hence, we study the impact of polling rates on the number of retransmissions. 
 Fig.~\ref{fig:retransmission_rate} shows the impact of different scenarios on the retransmission rate.
 Scenarios with 60~s polling intervals result in less retransmissions in comparison to 30~s scenarios. For example, in $UC1\;10\;OS\;30$ the retransmission rate is almost four times that of $UC1\;10\;OS\;60$. 
 
 Note that the $UC1\;5\;OS\;30$ and $UC1\;5\;OS\;60$ are not included in Fig.~\ref{fig:retransmission_rate}, because Use Case 1 requires eight or more outstations to have their binary operate packets inverted in order to generate a cascading failure in the Texas 2000-bus topology.
 
 

The number of DNP3 outstations polled also affects the amount of traffic the intruder is required to process. A larger number of outstations causes more traffic. The network buffer temporarily stores incoming packets before they are processed. Due to the limitation on the buffer size, the intruder may not be able to process all the traffic that traverses through it, which results in some attacks failing. Hence, the number of retransmissions increases. For example, from Fig.~\ref{fig:retransmission_rate} 
we can observe the retransmission rate for $UC2\;10\;OS\;30$ with ten polled outstations is almost 2.5 times that of the $UC2\;5\;OS\;30$ case with five polled outstations.


\subsection{Impact on Average Round Trip Time} 

\begin{figure}[b]
  \centering
\includegraphics[width=1\linewidth]{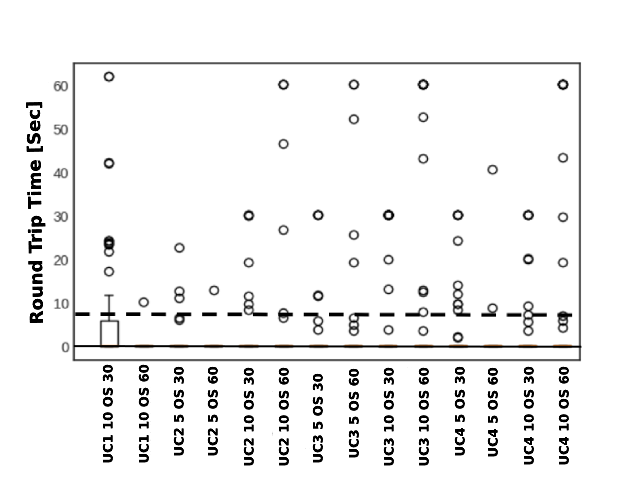}
  \caption{Round trip time for DNP3 traffic for each scenario.}
  \label{fig:dnp3_rtt}
  
\end{figure}

The RTT can be an indicator that an intruder is intercepting network traffic. RTT is also affected by the number of outstations polled and how often each is being polled. 
Fig.~\ref{fig:dnp3_rtt} shows the DNP3 packets that resulted in a high RTT and had to be retransmitted for different use cases.
The majority of the DNP3 traffic is received before the retransmission timer expires, 
shown by the dashed-line. The line shows the cut-off time for the DNP3 retransmission timer, which was set to 7.0~s.
We have observed that there is a longer delay when ten outstations are polled, compared with the number of retransmissions for the scenarios with five outstations. 

\subsection{Processing Time by Packet Type}\label{subsec:processingtime}
\begin{figure}[t]
  \centering
\includegraphics[width=1\linewidth]{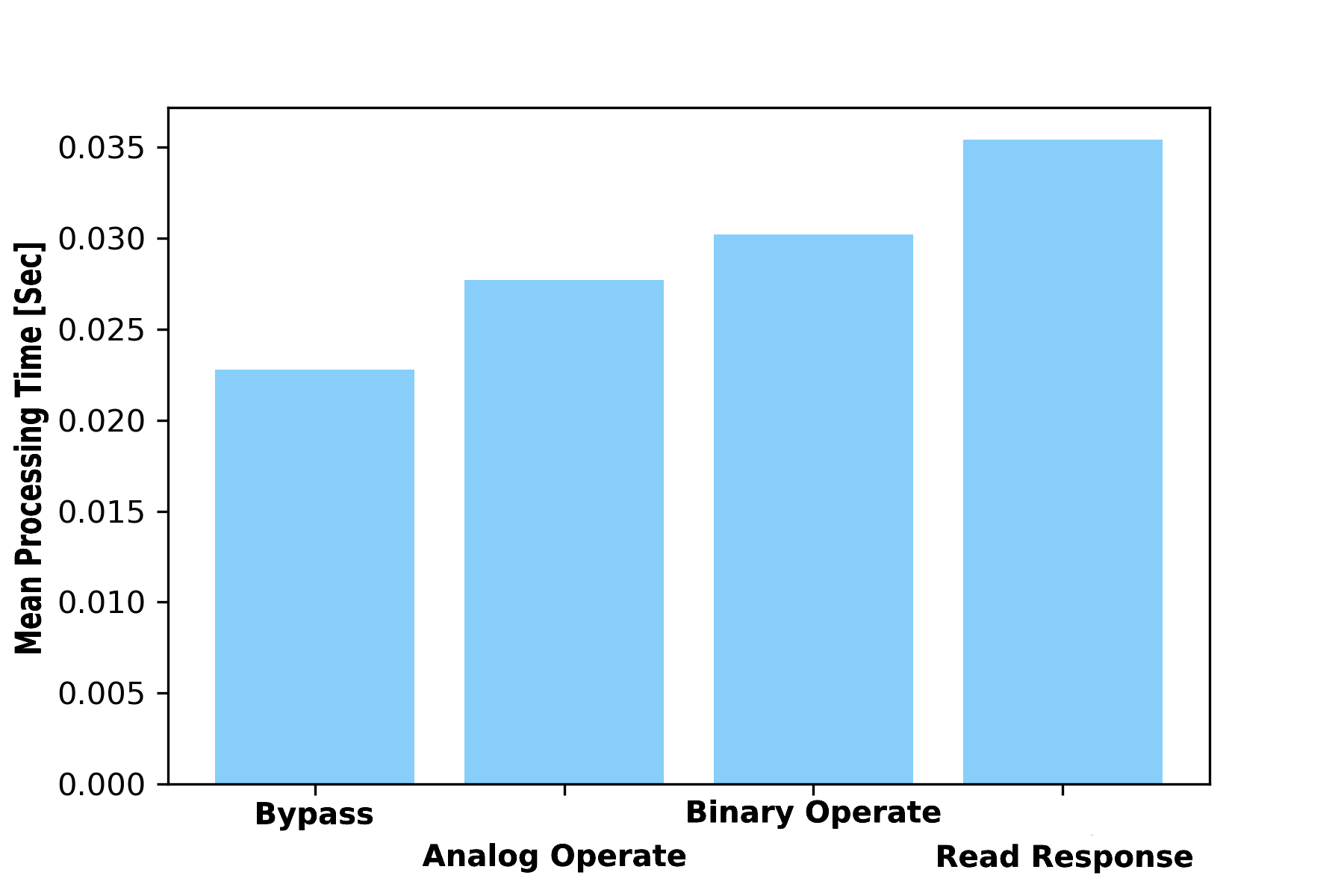}
  \caption{Average processing time by packet type at the adversary node.}
  \label{fig:dnp3_processingtime}
\end{figure}

The processing time at the adversary node is different for each packet type, as shown in Fig.~\ref{fig:dnp3_processingtime}, where we measure the average time to forward packets between the substation gateway and outstation by packet type. The lowest forwarding time and therefore most difficult packet type to detect the MiTM attack were the bypass packets, which took 22.775~ms to process. In the bypass packets, the adversary only modifies the source and destination MAC addresses of the frame. Next, the average processing time for an analog DIRECT OPERATE packet is higher at 27.693~ms, followed by binary DIRECT OPERATE taking an average of 30.217~ms. The highest processing time is for the read response packets, which take an average of 35.415~ms.

The processing time is directly correlated to the amount of traffic and the number of operations the MiTM script has to perform on each packet. For this reason, the bypass traffic took the least amount of time, as the adversary only updates the MAC address and forwards it to the original destination. Next, the operations to the binary and analog DIRECT OPERATE packets include modifying the value sent by the operator, recalculating the CRC, and forwarding the forged packet to the outstation. The read response packets take the most number of operations: two to three analog/binary values are modified, then the CRC for multiple data blocks 
are calculated and updated, and the packet is forwarded to the master.

\begin{figure}[t]
  \centering
\includegraphics[width=1\linewidth]{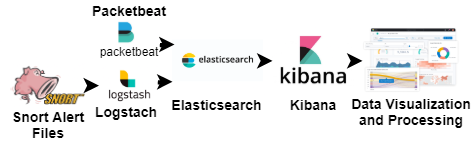}
  \caption{Data flow pipeline for Kibana data processing and graph generation.}
  \label{fig:experimental_model}
\end{figure}


\begin{figure*}[t]
  \centering
\includegraphics[width=1.0\linewidth]{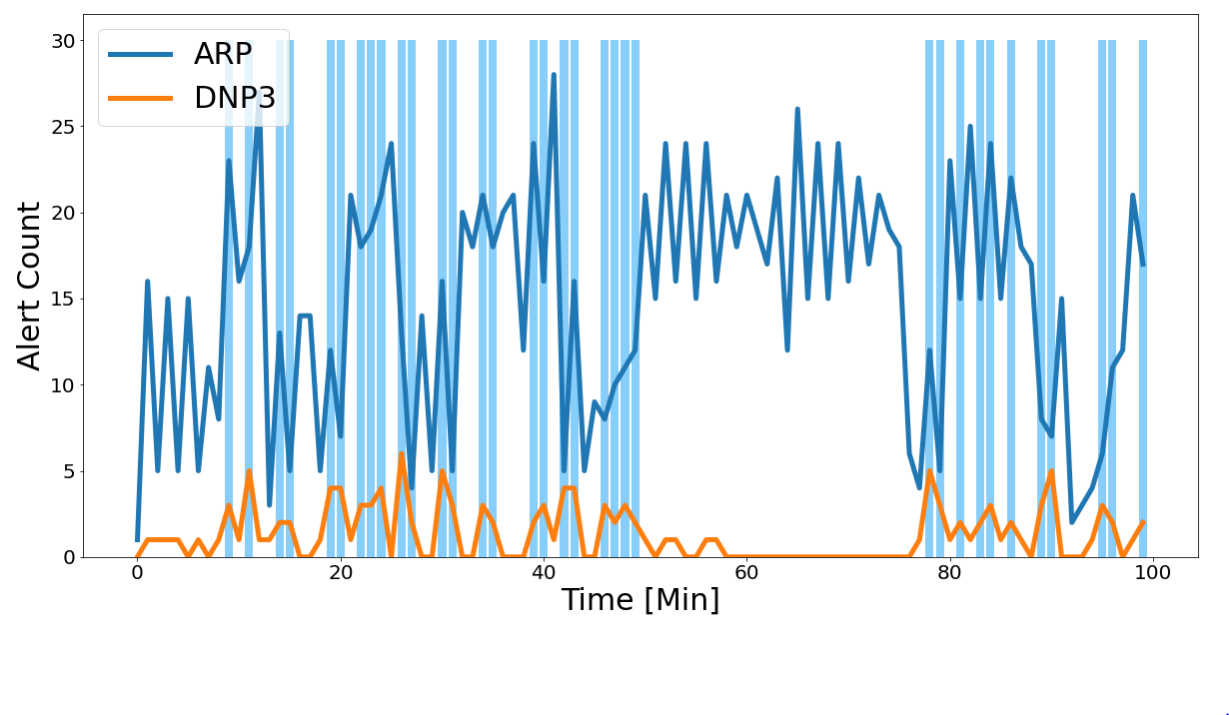}
\vspace{-1cm}
  \caption{ARP and DNP3 Snort alerts show potentially compromised DNP3 packets.}
  \label{fig:kibana_dashboard}
\end{figure*}

\subsection{Snort Detection}


In the RESLab testbed, the Snort logs and alerts are collected by Logstash, an open source tool that is used with Kibana to create dashboards for data analysis and visualization. After the Snort log data is processed by Kibana, the frequency of the ARP and DNP3 alerts show which DNP3 packets are being compromised at what time. 

During each use case, the PCAP data containing unsolicited ARP and DNP3 traffic, and the Snort's alert log files are collected by Logstash, which filters and formats the data. Then, as shown in Fig.~\ref{fig:experimental_model}, the data is stored in the Elasticsearch database. Kibana then acts as the front-end for the Elasticsearch database, by creating graphs that show the correlation between Snort alerts, ARP frames, and DNP3 packets.

To illustrate what is displayed in RESLab testbed's dashboard, Fig.~\ref{fig:kibana_dashboard} shows the ARP alerts that are generated when an ARP cache poisoning is detected based on the rules $R1$, $R2$, and $R3$, illustrated in Listing~\ref{snort_rules} in Section~\ref{detection}. In addition, DNP3 alerts are generated when DNP3 DIRECT OPERATE packets are detected as per $R4$ and $R5$ rules. 

We can conclude that during the time-period an ARP spoof is detected, the DNP3 DIRECT OPERATE packet is rerouted to the adversary's node, which indicates that the MiTM script is modifying the operation. By monitoring the RESLab's dashboard, it is possible to determine with a one-minute resolution which DNP3 packets are potentially compromised and should be discarded. The one-minute intervals that contain the DNP3 packets that should be discarded are shown by the blue lines in Fig.~\ref{fig:kibana_dashboard}.

%% file: Content/7_conclusion.tex
\section{Conclusion}
\label{conclusion}
Non-stealthy MiTM attacks can be developed and detected from features such as CRC mismatch, acknowledgements, and round trip times. However, if the intruder is stealthy enough to forge the CRCs, modify the acknowledgement packets, and reduce processing time by modifying selected DNP3 points in a payload, it can be difficult to detect such FDIs or FCIs.  

There are two main contributions this paper provides to the reader. The first is a step-by-step framework that describes how to implement four DNP3-based MiTM attacks on a SCADA system's network. The second is a method to detect MiTM attack traffic by correlating Snort IDS alerts with ARP and DNP3 packet data, using network metrics such as retransmission rate and average RTT. It is important to monitor these metrics to detect the signature of a MiTM attack. The processing time at the adversary causes the RTT to increase, and the increased RTT causes retransmissions. These causal behaviors can be extracted in the form of timestamped features for training machine-learning-based detection algorithms. 
%

Our results show that as the number of polled outstations increases, the DNP3 packets are delayed, and the efficiency of the MiTM attack decreases. This causes more DNP3 retransmissions, because consecutive packets from different outstations arrive at the adversary faster than the MiTM script can modify and forward the first packet. Also, we observe different processing times at the adversary for different types of DNP3 traffic. Read response packets had the longest processing time. Then, based on these results, we present defense recommendations such as showing how cyber telemetry can be used to detect stealthy MiTM attacks.  Results show that while rule-based IDS such as Snort can detect ARP spoofs using existing pre-processors, they can still result in higher false positives due to rule selection criteria. Hence, this work also provides recommendations for future work to incorporate metrics such as average RTT and retransmission rate into security-centric data analysis on anomaly-based attack detection. 

%